\documentclass[aps,prl,twocolumn,amsmath,amssymb]{revtex4}

\usepackage{epsfig}
\usepackage{graphicx}
\usepackage{amsfonts}
\usepackage{amssymb}
\usepackage{amsmath}
\usepackage{bm}

\renewcommand{\vec}[1]{{\mathbf #1}}

\bibpunct{(}{)}{;}{s}{,}{,}

\begin{document}

\title{Voltage Induced Dynamical Quantum Phase Transitions in Exciton Condensates}

\author{Moon Jip Park}
\affiliation{Department of Physics, University of Illinois, Urbana, Il, 61801}
\altaffiliation{Micro and Nanotechnology Laboratory, University of Illinois, Urbana, Il 61801}

\author{E. M. Hankiewicz}
\affiliation{Institut f\"{u}r Theoretische Physik und Astrophysik,
Universit\"{a}t W\"{u}rzburg, Am Hubland, 97074 W\"{u}rzburg, Germany}

\author{Matthew J. Gilbert}
\affiliation{Department of Electrical and Computer Engineering, University of Illinois, Urbana, Il, 61801}
\altaffiliation{Micro and Nanotechnology Laboratory, University of Illinois, Urbana, Il 61801}

\date{\today}

\begin{abstract}
We explore non-analytic quantum phase dynamics of dipolar exciton condensates formed in a system of 1D quantum layers subjected to voltage quenches. We map the exciton condensate physics on to the pseudospin ferromagnet model showing an additional oscillatory metastable and paramagnetic phase beyond the well-known ferromagnetic phase by utilizing a time-dependent, non-perturbative theoretical model. We explain the coherent phase of the exciton condensate in quantum Hall bilayers, observed for currents equal to and slightly larger than the critical current, as a stable time-dependent phase characterized by persistent charged meron flow in each of the individual layers with a characteristic AC Josephson frequency.  As the magnitude of the voltage quench is further increased, we find that the time-dependent current oscillations associated with the charged meron flow decay, resulting in a transient pseudospin paramagnet phase characterized by partially coherent charge transfer between layers, before the state relaxes to incoherent charge transfer between the layers.
\end{abstract}
\maketitle

The dipolar exciton condensate (DEC), which can be directly translated into the pseudospin ferromagnetism model (PFM)\cite{MacDonald2001}, have provided dramatic observations of collective phenomena in a broad swath of host systems including: cold atoms \cite{Baranov2002,Neely2010,Hadzibabic2006, Potter2010} semiconductor microcavities \cite{Balili2007, Christopoulos2007, Butov2002}, and semiconductor quantum wells \cite{Kellog2004,Tutuc2004, Snoke2002, Tiemann2008, Yoon2010, Sinclair2011, Nandi2013}. In each of these settings, the Coulomb interaction between spatially segregated charge carriers drives many-body phase transition from the normal Fermi liquid phase to that of a superfluid. Beyond the interesting correlated physics these systems demonstrate, they continue to harbor tantalizing prospects for ultra-efficient, electrically-tunable information processing systems based on predictions of elevated Kosterlitz-Thouless transition temperatures ($T_c$) without the need for external magnetic fields to quench the kinetic energy\cite{Dellabetta2013}. These prospects may be directly traced to the realization of new Dirac material systems such as graphene \cite{Min2008,Gilbert2009,Gilbert2010} and time-reversal invariant topological insulators \cite{Seradjeh2009,Kim2012, Cho2011, Tilahun2011,Budich2014}. In particular, recent experimental work in monolayers of graphene separated by hexagonal boron nitride show signatures of correlated behavior well-above cryogenic temperatures \cite{Gorbachev2012}.

Of the signatures indicative of the collective phenomena associated with PFM, some of the most dramatic are those found in carrier transport. Within the context of carrier transport, one of the most fundamental parameters is the critical current ($I_c$), the maximum current that the DEC can sustain by simply reorganizing its order parameter. The behavior of the PFM is well-understood below $I_c$ where the system exhibits coherent superfluid flow, characterized by time-independent coherent current flow and perfect Coulomb drag \cite{Eisenstein2004,Su2008}. However, in the region past the critical current, there is a clear deficiency concerning PFM system behavior as voltage quenches resulting in current flow greater than $I_c$ are applied. Naturally, in this regime, linear response approach is not be applicable and non-perturbative approaches are required. As a corollary, recent study in dynamical phase transitions in transverse field Ising model have shown non-analytic behavior when considering real-time quenches from ferromagnet to paramagnet\cite{Heyl2013} whose behavior is not captured within framework of linear response theory.

Here, we theoretically explore the behavior of a generic PFM system beyond linear response theory. We consider spatially segregated 1D semiconducting layers (though our conclusions are valid for 2D systems) using a time-dependent Kadanoff-Baym (TDKB) formalism \cite{Myohanen2009, Stefanucci2013} subjected to time-dependent voltage quenches. We are motivated by recent experiments on DEC \cite{Yoon2010,Nandi2013} where, surprisingly, at an interlayer voltage equal to the critical voltage, V$_c$, the condensate behaves in a manner consistent with the fully coherent regime. We explain this observation as a voltage-driven competition between the PFM and a pseudospin paramagnetic (PPM) phase characterized by a time-dependent coherent exciton state which recovers its coherence by periodically launching merons with charge $\frac{q}{2}$. This new regime could serve as an ideal setting for a direct measurement of the quantized charge associated with merons, which should be more definitive than observations of non-zero longitudinal resistance of condensates at finite temperatures\cite{Kellog2004,Tutuc2004} or indirect influence of topological excitations on Shapiro steps \cite{Hyart2013}. As the magnitude of the voltage quenches are increased well-beyond $V_{c}$, we find that the system can no longer relax the superfluid flow by inducing merons and the interlayer coherence, which characterizes the PFM phase, is lost and the layers behave independently, as expected from experiments \cite{Yoon2010,Nandi2013}.  Moreover, our analysis shows that the condensate in the crossover regime not only will respond to microwave frequencies \cite{Hyart2013} but shows new possibilities as a voltage-tunable electrical oscillator.
\begin{figure}
\includegraphics[width=0.4\textwidth]{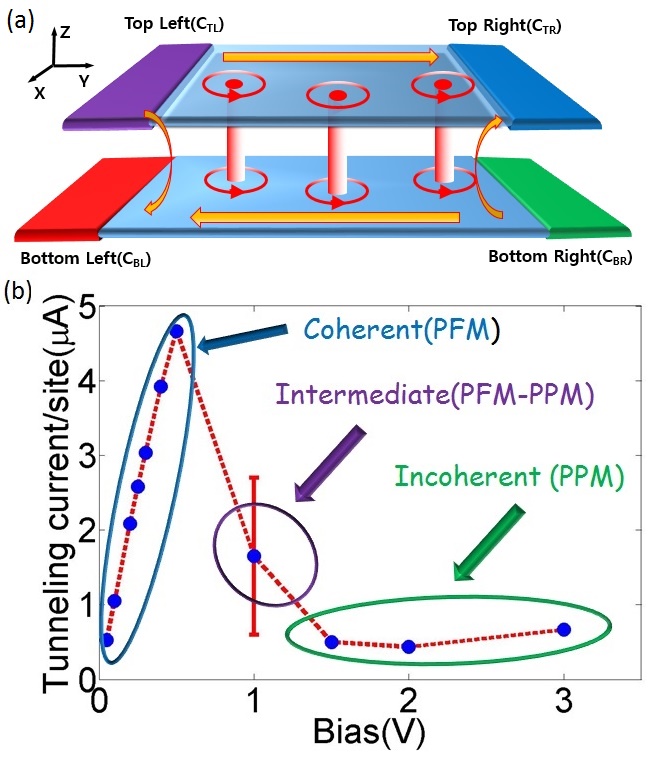}
\caption{\label{fig:FIG1} (a) Schematic illustration of a pseudospin ferromagnet with contacts attached to each of the edges of the system. The arrows indicate the directions of the inter and interlayer quasiparticle motion in each layer. Above $I_{c}$ the system proliferates charged vortices which propagate in the same direction within each layer. (b) Plot of the calculated time-averaged coherent tunneling current from $C_{TL}$ to $C_{BL}$  ($C_{BR}$ to $C_{TR}$) as a function of interlayer bias. We obtain the experimentally expected behavior which may be associated with the existence of three distinct pseudospin regimes: coherent (PFM), intermediate (PFM-PPM), and incoherent (PPM)}
\end{figure}

We begin in Fig.~\ref{fig:FIG1}(a) where we schematically picture the system of interest. Here we have two 1D semiconducting layers in which the top layer is assumed to contain electrons and the bottom layer is assumed to contain an equal population of holes. For simplicity, we assume that the layers are free from disorder. We attach contacts to the left and right ends of the top layer ($C_{TL}$ and $C_{TR}$) and the bottom layer ($C_{BL}$ and $C_{BR}$) from which we inject and extract currents. With the system defined, we may now write the tight-binding Hamiltonian for the system as
\begin{equation}
\label{Eq:Hdef}
\bold{H}=\begin{bmatrix}H_{top} & 0 \\ 0 &H_{bot}\end{bmatrix} + \sum_{\mu = x,y,z} \hat{\mu} \cdot \vec{\Delta} \otimes \sigma_\mu .
\end{equation}
The first term on the right hand side of Eq.(\ref{Eq:Hdef}) is the single-particle non-interacting term while the second term is a mean-field interaction term considered here to be purely local in nature. In Eq. (\ref{Eq:Hdef}), $H_{top}$ and $H_{bot}$ are the simple 1D single subband effective mass chains with hopping energy of $t_x$ = $\pm$3 eV.  The interaction term in Eq. (\ref{Eq:Hdef})  includes the Kronecker product of  pseudospin effective ferromagnetic bonding, $\Delta$,  with Pauli spin matrices $\sigma_{\mu}$ and has the form \cite{MacDonald2001, Burkov2002, Gilbert2010, Kim2012b}
\begin{equation}
\label{Eq:deltadef}
\bold{\Delta}=(\Delta_{sas} + U \bold{m}_{ps}^x) \, \hat{x}+ U \bold{m}_{ps}^y \, \hat{y}
\end{equation}
where $\Delta_{sas}$ is the single particle tunneling amplitude, $U$ is the strength of the on-site electron-electron interactions, and the pseudospin-magnetization $\bold{m_{ps}} = \frac{1}{2}\operatorname{Tr}[\rho_{ps}\bm{\sigma}]$.
 $\rho_{ps}$ is the $2\times2$ Hermitian pseudospin density matrix which we define as,
\begin{equation}
\rho_{ps}=\begin{bmatrix} \rho_{\uparrow\uparrow}&\rho_{\uparrow\downarrow}\\\rho_{\downarrow\uparrow}&\rho_{\downarrow\downarrow} \end{bmatrix}.
\end{equation}
The diagonal terms of pseudospin density matrix (\(\rho_{\uparrow\uparrow}, \rho_{\downarrow\downarrow}\)) are the electron densities of top and bottom layers. In Eq. (\ref{Eq:deltadef}) , we justify the omission of the exchange potential in the $\hat{z}$ direction because this contribution is dominated by the electric potential difference between layers induced by the interlayer bias voltage quench. From the definition of $m_{ps}$, we obtain directional pseudospin-magnetization, which define the magnitude of the pseudospin order parameter, and the planar pseudospin angle, $\phi_{ps}$, which corresponds physically to the phase difference between quasiparticles in the two layers. We have set $U$ = -5 eV and $\Delta_{sas}$ = $10^{-3}$eV  in Eq. (\ref{Eq:deltadef}) . Once the parameters are selected, the system is self-consistently iterated until convergence of the order parameter is reached between successive iterations of the density matrix. Our choices of parameters result in a gap size of $\Delta$ = 0.53eV, and a coherence length, $\hbar v_{F} / \Delta$, of 11 lattice points. The choice of the parameters does not hurt qualitative physics we aim to address.

In order to incorporate the time-dependent dynamics associated with voltage quenches of the PFM, we must solve the Kadanoff-Baym equation \cite{Stefanucci2013, Stan2009} using the self-consistently obtained Hamiltonian for the PFM at $t=0$ as a starting point,
\begin{equation}
\label{Eq:kbe}
(i\partial_t-\bold{H})G(t,t')=\delta(t,t')+\int{dt_1}{\Sigma(t,t_1)G(t_1,t')}.
\end{equation}
In Eq. (\ref{Eq:kbe}), $G$ is the Green's function, $G_{ij}^<(t)=i\langle c_j^\dagger c_i \rangle= i\rho_{ji}(t)$ \cite{Perfetto2010}, with $\rho$ as the single particle density matrix, and $\Sigma(t,t')$ as the self-energy term. We may significantly reduce the complexity of the time propagation when the interactions are local in time. In this case, the off-diagonal terms in the self-energy must vanish resulting in a very simple expression for the self-energy, $\Sigma_{ij}(t,t')=\delta(t-t')v_{ji}G_{ij}^<(t,t')$ which includes the exchange interaction, $v_{ij}$.

With the methodology established, we now apply positive voltage to the top left contact ($V_{TL} = V_{int}$) for times $t>0$ and  examine the current flow into and out of each respective contact. In this voltage configuration, the stability of exciton condensate is established by global, time-dependent phase transformation. As we are interested in voltage based phase transitions, we may delineate these phases with the definition of the critical voltage, $V_{c}$, or the interlayer voltage which produces $I_{c}$. In Fig.~\ref{fig:FIG1}(b), we plot the time averaged interlayer current magnitude flowing from $C_{TL}$ to the $C_{BL}$ as a function of the bias applied to $V_{TL}$.

We immediately notice Fig. ~\ref{fig:FIG1}(b) can be directly compared with the known experimental interlayer transport properties over the entire range of interlayer voltages \cite{Yoon2010}.  Specifically, we recover the observed experimental trends in steady-state interlayer conductivity in PFM systems for  $V_{TL}-V_{c} < 0$ and $\phi_{ps} \neq 0$, corresponding to the growth of the coherent tunnel current of the exciton condensate with the applied voltage. In this case, the interlayer current
\cite{Kim2012b} is
\begin{equation}
 \label{eq:current}
J_{int}(r)=i[H,N_{top}]=t_{TB}c_{T}^\dagger c_{B}-t_{BT}c_{B}^\dagger c_{T}=2\Delta_{sas}m^y_{ps}
\end{equation}
where $t_{TB}$ is effective interlayer hopping from top to bottom and $c_{T(B)}$ is a quasiparticle annihilation operator at top(bottom). In the PFM regime, when current is injected from $C_{TL}$ ($C_{TR}$), an equal and opposite amount of current will flow into $C_{BL}$ ($C_{BR}$).
This perfect Coulomb drag may be understood from a simple analogy to Andreev reflections in superconductivity\cite{Su2008,Kim2012b,Gilbert2010}. Within the PFM regime, it is always possible to obtain a self-consistent steady state solution between the equations of motion, the electrostatics, and the interactions with respect to global time-dependent phase rotation. In other words, the static limit of Landau-Lifshitz-Slonczewski (LLS) equation must posses a solution\cite{Su2010}.
\begin{figure}
\includegraphics[width=0.5\textwidth]{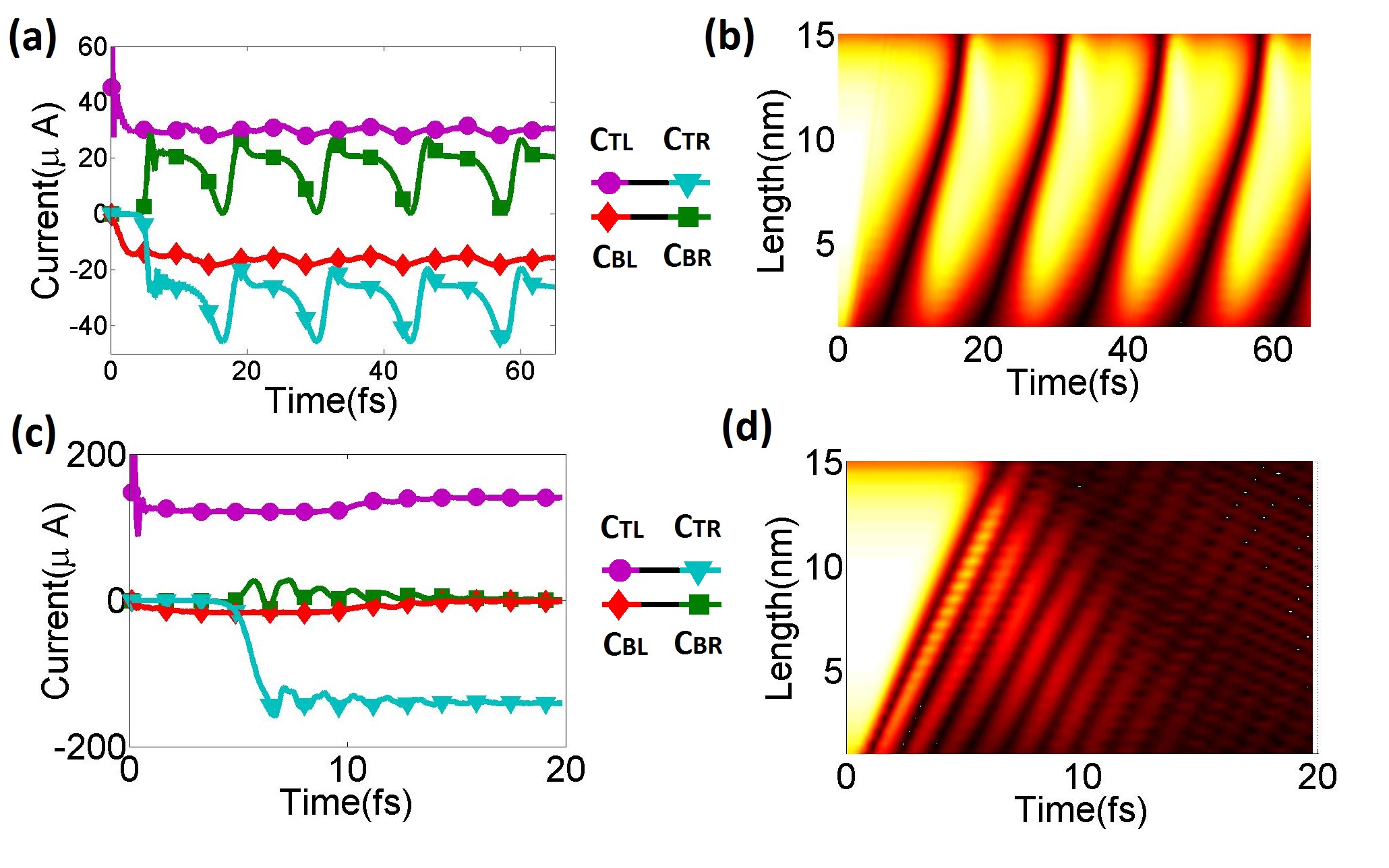}
\caption{\label{fig:FIG2} (a) Plot of the terminal currents versus time for $V_{TL}=1.2~V$ for the layers containing 50 points in the computational space. (b) Plot of the magnitude of the pseudospin order parameter as a function of time and length.  (c) Plot of the terminal currents versus time for $V_{TL}=4.0~V$. (d) Plot of the magnitude of the pseudospin order parameter as a function of time and length.}
\end{figure}

When the applied voltage is equivalent to the critical voltage, $V_{TL}-V_{c} \approx 0$, the interlayer current reaches $I_c$ and we observe an abrupt drop in the magnitude of the interlayer current transfer in Fig. \ref{fig:FIG1} along with a suppression of the interlayer Coulomb drag. This drop signals the termination of the purely PFM regime and the onset of an intermediate metastable regime. While the drop in interlayer charge transfer is expected based on the misalignment of the layer Fermi surfaces, the large error bars for tunneling current indicate the presence of significant oscillations in the terminal currents. This behavior is associated with the persistent launching of merons in both layers which slows down the condensate velocity and recovers the coherent phase. As will be discussed more below, this persistent meron launching explains the recovery of a PFM-like state in quantum Hall bilayers past $I_c$. Further increase in the applied potential in which $V_{TL} - V_{c} \gtrsim \Delta$ show the magnitude of the interlayer current continues to decrease as the two layers become increasingly energetically separated. In this range of voltages, the system is in the incoherent or PPM phase in which the magnitude of the interlayer current appears to be governed solely by the value of $\Delta_{sas}$, in agreement with previous experimental results.\cite{Yoon2010,Nandi2013}.

To form a more complete understanding of the nature of the terminal currents past $V_{c}$, we examine the resulting terminal currents and $|m_{ps}|$ for several interlayer voltages each resulting in $V_{TL} > V_{c}$. In Fig. \ref{fig:FIG2}(a), we apply a bias of $V_{TL} = 1.2~V$ that results in a current within the metastable regime. Indeed, in Fig. \ref{fig:FIG2}(a), we see that each of the terminal currents begins to stably oscillate with the largest magnitude oscillations appearing in $I_{TR}$ and $I_{BR}$. These oscillations are signatures of a competition between the PFM and PPM phases with a frequency consistent with the AC Josephson frequency proportional to $e(V-V_c) / h$\cite{Notarys1971,Leggett2001}. Its maximum value is limited by excitonic gap size, using experimentally measured value of gap, which corresponds to a frequency of $16.7$ GHz\cite{Giudici2010} (See Supplement). The coherence between the layers triggers an electron current and equivalent hole current flow. Time-averaged current flow in bottom layer is lower than top layer as a a consequence of the partial suppression of coherence brought about by the competition between the two distinct phases and non-zero spatial overlap between successive merons. At minimum points of $I_{BR}$ and $I_{TR}$, the layers temporarily lose coherence. The loss of coherence results in $I_{BR}$ possessing nearly zero value and a peak in $I_{TR}$ indicating that the observed behavior is associated with the negative density fluctuations.

Beyond $V_c$, we expect there is non-zero electric field inside the system approximately given by $E\approx(V_{TL}-V_c)/L$ that accelerates the exciton pairs across the system. By launching the vortex, the phase gradient is reduced and the exciton pairs in the PFM phase are decelerated in order to keep a constant superfluid velocity. To be more specific, when the $\phi_{ps}$ winds into the $\hat{z}$-direction, the system launches a electron-like meron, which has $\frac{q}{2}$ charges at each layer, that flow from the left of the system to the right at an applied bias of $V_{TL}=1.2~V$, as seen in Fig. \ref{fig:FIG2}(b). The merons are topological defects \cite{Girvin1997,Moon1995} that break the order of the condensate and retain with them pseudospin order that points solely in the $\hat{z}$-direction corresponding to zeros in $|m_{ps,xy}|$ accompanied by a $\pi$ phase slip in the condensate after which coherence is restored. Therefore, the stable oscillations in the terminal currents are attributed to voltage-driven fluctuations between the PFM and PPM phases characterized by the entering and exiting of meron pair from the contacts (See Supplement). The  maxima in $I_{BR}$ and $I_{TR}$ indicate that the coherence is recovered after the fluctuation passes. In Fig. \ref{fig:FIG2}(a), the condensate is not fully recovered at maximum points, as the meron bound states are not fully localized and the non-zero spatial overlap between bound states forms the discrepancy from strong interaction limit (See Supplement).
\begin{figure}
\includegraphics[width=0.45\textwidth]{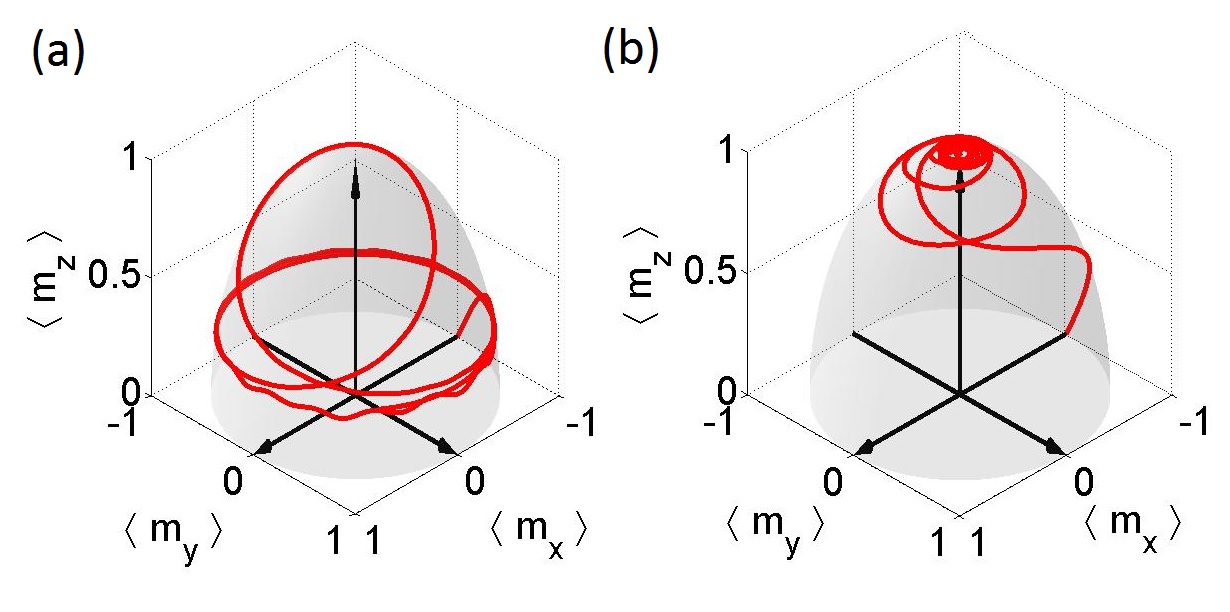}
\caption{\label{fig:FIG4} Time evolution of pseudospin orientation at x = 40 lattice point for (a) the intermediate PFM-PPM metastable oscillations at an applied bias of $V_{TL}=1.2~V$ (b) the PPM phase at an applied bias of $V_{TL}=4.0~V$. Each of these figures is taken within the time frame of one phase slip period for visual simplicity. }
\end{figure}

In Fig. \ref{fig:FIG2}(c), we see another transition from the intermediate metastable oscillations between PFM and PPM phases to a stable PPM phase, which arises when $m_z$ dominates the pseudospin orientation. In this regime, the bias induced energy separation between the two layers wins a competition with the coherence of bilayer. As a result, in PPM phase, the current flows from $C_{TL}$ to $C_{TR}$ with only a transient response in $C_{BL}$ and $C_{BR}$. Yet within the transient regime the current flowing to the lead $C_{BR}$ is positive indicating the presence of transient interlayer coherence in the system. Fig. \ref{fig:FIG2}(d) shows the exponential decay of order parameter magnitude as the exchange enhancement is lost and the value asymptotes towards the non-interacting $\Delta_{sas}$ with $\phi_{ps}$ pointing in the $\hat{z}$-direction. It is critical to note that, in closed system, the transition to PPM phase is forbidden since total magnetic moment in $\hat{z}$-direction $m_{z-tot}=\sum_i m_z$ is a roughly conserved quantity within time scale $1/\Delta_{sas}$. However, open contacts act as a thermal reservoir that exchanges both energy and pseudospin. Thus, the existence of the reservoir allows the thermalization to PPM state. In other words, at $t>0$, direct insertion and extraction of pseudospin (quasiparticles) through open contact can relax the system to the PPM phase.

To more clearly illustrate voltage induced phase transition, Fig. \ref{fig:FIG4} shows trajectory of normalized pseudospin evolution along the Bloch sphere. Fig. \ref{fig:FIG4}(a) shows oscillatory behavior between PFM-PPM phase characteristic of the metastable phase as the pesudospin orientation precesses in $x-y$ plane. It precesses out of plane to touch $z$-axis before returning to $x-y$ plane when launched meron pass through the observation point. After the orientation returns to x-y plane, it persists precession in its orbit until the next meron reaches the observation point. The precession in $x-y$ plane is a consequence of global phase evolution and the acceleration of the superfluid. When the pseudospin phase touches the north pole of the pseudospin Bloch sphere, it winds once about the pole as a direct reflection of presence of the meron. Regress of pseudospin to $x-y$ plane indicates the recovery of phase coherence. In Fig. \ref{fig:FIG4}(b), the excessive bias breaks the coherence between the layers forcing the transition from PFM to PPM phase. After the initial transient behavior, the pseudospin phase angle eventually precesses into the $z$-direction, consistent with the current-induced phase transition to the PPM phase. In transient regime before the PPM phase is fully established, pseudospin winds north pole a few times before reaching its stable out-of-plane orientation along the pseudospin Bloch sphere and confirming quenched phase transition.

In conclusion, we have studied the non-equilibrium time-dependent dynamics of PFM phases focusing on the behavior past $V_c$. For voltages $V_{TL} - V_{c} < 0$, the system exhibits a PFM denoted by perfect drag counterflow between the two layers. For increased interlayer voltages $V_{TL} - V_c \approx 0$, the system exhibits stable oscillation between the PFM phase and the PPM phase characterized by oscillations in terminal currents corresponding to the continuous launching of charge $\frac{q}{2}$ merons across the superfluid. The presence of oscillatory phase and the recovery of the PFM help to explain experimental observations past $I_{c}$ in quantum Hall bilayer. When the interlayer bias exceeds, $V_{TL} - V_{c} \gtrsim \Delta$ the coherence is destroyed and the system transitions into the PPM phase.

\section{Acknowledgements}
E.M.H. thanks German Research Foundation (DFG) under grant HA 5893/4-1 within SPP 1666 for a financial support and the ENB graduate school ”Topological insulators". M.J.P. and M.J.G. acknowledge financial support from the Office of Naval Research (ONR) under grant N0014-11-1-0123 and the National Science Foundation (NSF) under grant CAREER EECS-1351871. M.J.P. acknowledge useful discussion from Gil Young Cho and Brian Dellabetta.

\begin{acknowledgements}
\end{acknowledgements}


\begin{thebibliography}{}

\bibitem{MacDonald2001}
A. H. MacDonald,
Physica B {\bf 298}, 129 (2001).

\bibitem{Baranov2002}
M. A. Baranov, M. S. Mar'enko, V. S. Rychkov, and G. V. Shlyapnikov,
Phys.\ Rev.\ A {\bf 66}, 013606 (2002).

\bibitem{Hadzibabic2006}
Z. Hadzibabic, P. Kr\"uger, M. Cheneau, B. Battelier, and J. Dalibard,
Nature (London) {\bf 441}, 1118 (2006).

\bibitem{Neely2010}
T. W. Neely, E. C. Samson, A. S. Bradley, M. J. Davis, and B. P. Anderson,
Phys.\ Rev.\ Lett.\  {\bf 104}, 160401 (2010).

\bibitem{Potter2010}
A. C. Potter, E. Berg, D. W. Wang, B. I. Halperin, and E. Demler,
Phys.\ Rev.\ Lett. {\bf 105}, 220406 (2010).

\bibitem{Balili2007}
R. B. Balili, V. Hartwell, D. Snoke, L. Pfeiffer, and K. West,
Science {\bf 316}, 1007 (2007).

\bibitem{Christopoulos2007}
S. Christopoulos, G. B. H\"orger von H\"ogersthal, A. J. D. Grundy,
P. G. Lagoudakis, A. V. Kavokin, J. J. Baumberg, G. Christmann,
R. Butt\'e, E. Feltin, J. F. Carlin and N. Grandjean
Phys.\ Rev.\ Lett.\  {\bf 98}, 126405 (2007).

\bibitem{Butov2002}
L. V. Butov, A. C. Gossard, and D. S. Chemla, Nature (London), {\bf 418}, 751 (2002).

\bibitem{Yoon2010}
Y. Yoon, L. Tiemann, S. Schmult, W. Dietsche, K. von Klitzing,
and W. Wegscheider, Phys.\ Rev.\ Lett.\  {\bf 104}, 116802 (2010).

\bibitem{Nandi2013}
D. Nandi, T. Khaire, A. D. K. Finck, J. P. Eisenstein, L. N. Pfeiffer and K. W. West,
Phys.\ Rev.\ B {\bf 88}, 165308 (2013).

\bibitem{Kellog2004}
M. Kellogg, J. P. Eisenstein, L. N. Pfeiffer, and K. W. West,
Phys.\ Rev.\ Lett.\ {\bf 93}, 036801 (2004).

\bibitem{Tutuc2004}
E. Tutuc, M. Shayegan, and D. A. Huse,
Phys.\ Rev.\ Lett.\  {\bf 93}, 036802 (2004).

\bibitem{Snoke2002}
D. Snoke, S. Denev, Y. Liu, L. Pfeiffer, and K. West,
Nature (London) {\bf 418}, 754 (2002).

\bibitem{Tiemann2008}
L. Tiemann, W. Dietsche, M. Hauser, and K. von Klitzing,
New J. Phys. {\bf 10}, 045018 (2008).

\bibitem{Sinclair2011}
N. W. Sinclair, J. K. Wuenschell, Z. V\"or\"os, B. Nelsen, D. W. Snoke,
M. H. Szymanska, A. Chin, J. Keeling, L. N. Pfeiffer, and K. W. West,
Phys.\ Rev.\ B {\bf 83}, 245304 (2011).

\bibitem{Dellabetta2013}
B. Dellabetta and M. J. Gilbert,
J. \ Comp. \ Electron. {\bf 12}, 248 (2013).

\bibitem{Min2008}
 H. Min, J. J. Su, and A. H. MacDonald,
 Phys.\ Rev.\ B.\ {\bf 78} 121401 (2008).

\bibitem{Gilbert2009}
M. J. Gilbert and J. Shumway, J.\ Comput.\ Electron.\ {\bf 8}, 51 (2009).

\bibitem{Gilbert2010}
M. J. Gilbert, Phys.\ Rev.\ B.\ {\bf 82} 165408 (2010).

\bibitem{Seradjeh2009}
 B. Seradjeh, J.E. Moore, and M. Franz,
 Phys.\ Rev.\ Lett.\ {\bf 103} 066402 (2009).

\bibitem{Kim2012}
Y. Kim, E. M. Hankiewicz and M. J. Gilbert, Phys.\ Rev.\ B {\bf 86}, 184504 (2012).

\bibitem{Cho2011}
 G. Y. Cho, and J. E. Moore,
 Phys.\ Rev.\ B.\ {\bf 84} 165101 (2011).

\bibitem{Tilahun2011}
D. Tilahun, B. Lee, E. M. Hankiewicz, and A. H. MacDonald,
Phys.\ Rev.\ Lett.\ {\bf 107}, 246401 (2011).

\bibitem{Budich2014}
J. C. Budich, B. Trauzettel, and P. Michetti Phys.\ Rev.\ Lett.\ {\bf 112}, 146405 (2014).

\bibitem{Gorbachev2012}
 R. V. Gorbachev, A. K. Geim, M. I. Katsnelson, K. S. Novoselov, T. Tudorovskiy,	 I. V. Grigorieva,	A. H. MacDonald, S. V. Morozov,	K. Watanabe, T. Taniguchi	
 and L. A. Ponomarenko
Nat.\ Phys.\  {\bf 8},  896 (2012).

 \bibitem{Eisenstein2004}
 J. P. Eisenstein, and A. H. MacDonald,
 Nature {\bf 432}, 691 (2004).

 \bibitem{Su2008}
 J. J. Su, and A. H. MacDonald,
 Nature.\ Phys.\ {\bf 4} 799-802 (2008).

 \bibitem{Heyl2013}
 M. Heyl, A. Polkovnikov, and S. Kehrein
Phys.\ Rev.\ Lett.\ {\bf 110} 135704 (2013).

\bibitem{Hyart2013}
T. Hyart and B. Rosenov,
Phys.\ Rev.\ Lett.\ {\bf 110}, 076806 (2013).
 \bibitem{Myohanen2009}
P. Myohanen, A. Stan, G. Stefanucci, and R. V. Leeuwen,
 Phys.\ Rev.\ B.\ {\bf 80} 115107 (2009).

\bibitem{Stefanucci2013}
G. Stefanucci, R. van Leeuwen {\em Nonequilibrium Many-Body Theory of Quantum Systems},
Cambridge University Press (2013).

\bibitem{Ozyuzer2009}
L. Ozyuzer, Y. Simsek, H. Koseoglu, F. Turkoglu, C. Kurter, U. Welp, A. E. Koshelev, K. E. Gray, W. K. Kwok, T. Yamamoto, K. Kadowaki, Y. Koval, H. B. Wang and P. M\"uller
Supercond.\ Sci.\ Technol.\  {\bf 22}, 114009 (2009).

\bibitem{Burkov2002}
A. A. Burkov and A. H. MacDonald
Phys.\ Rev.\ B, {\bf 66}, 115320 (2002).

 \bibitem{Kim2012b}
Y. Kim, A. H. MacDonald, and M. J. Gilbert,
 Phys.\ Rev.\ B.\ {\bf 82} 165424 (2012).

\bibitem{Stan2009}
A. Stan, N. E. Dahlen, and R. V. Leeuwen,
J.\ Chem.\ Phys.\ {\bf 130} 224101 (2009).

 \bibitem{Perfetto2010}
E. Perfetto, G. Stefanucci, and M. Cini
 Phys.\ Rev.\ B.\ {\bf 82} 035446 (2010).

 \bibitem{Su2010}
 J. J. Su, and A. H. MacDonald,
 Phys.\ Rev.\ B {\bf 81} 184523 (2010).

\bibitem{Girvin1997}
For a review, S. M. Girvin and A. H. MacDonald, in
{\em Perspectives in Quantum Hall Effects}
,edited by A. Pinczuk and S. Das Sarma (Wiley, New York,
1997)

\bibitem{Moon1995}
K. Moon, H. Mori, K. Yang, S. M. Girvin, A. H. MacDonald, L. Zheng, D. Yoshioka, and S.-C. Zhang
Phys.\ Rev.\ B  {\bf 51}, 5138 (1995).

\bibitem{Notarys1971}
H. A. Notarys and J. E. Mercereau,
 Physica {\bf 55} 424 (1971).

\bibitem{Leggett2001}
A. J. Leggett,
 Rev.\ Mod.\ Phys.\ {\bf 73} 307 (2001).

 \bibitem{Giudici2010}
 P. Giudici, K. Muraki, N. Kumada, and T. Fujisawa
 Phys.\ Rev. \ Lett. {\bf 104}, 056802 (2010).


\end{thebibliography}

\end{document}


\appendix
\appendix
\section{Condensate Recovery}
\begin{figure}
\includegraphics[width=0.45\textwidth]{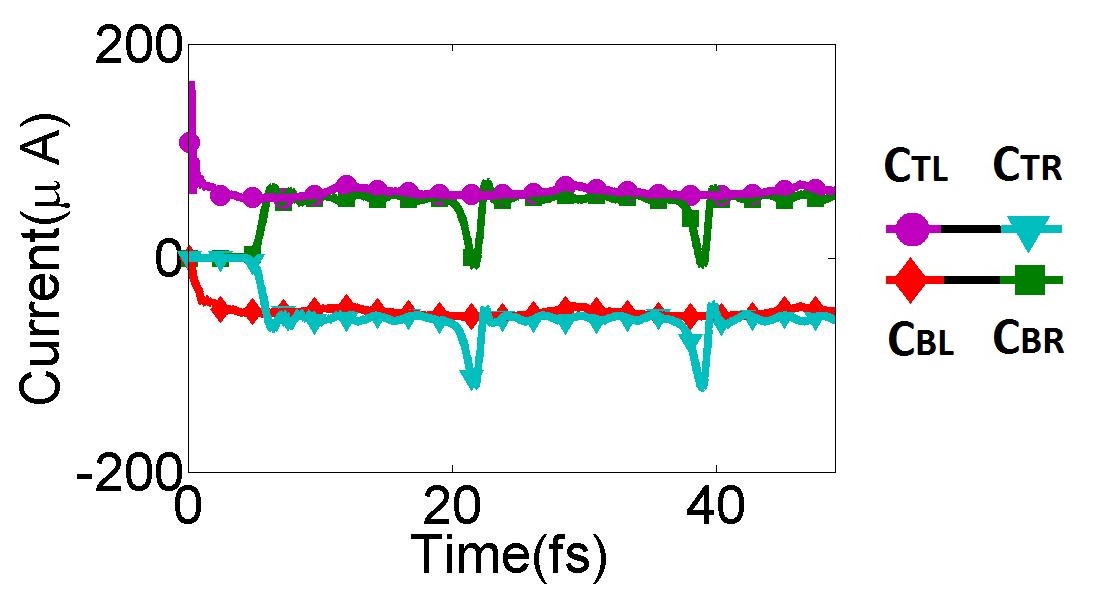}
\caption{\label{fig:fullrec} Current flow at each contact in oscillatory phase at $U=-7~eV$ and $V_{TL}=3~V$. In high $U$ limit, the charged vortex states are well-localized and allow for the full recovery of interlayer coherence after each oscillation.}
\end{figure}

In the main text of the paper, we have asserted that after the merons pass from the system into the contacts, then the system recovers full interlayer coherence when the system is in the PFM-PPM oscillatory phase. This is difficult to observe in the main text as the interaction strength of $U=-5~eV$ which led to a long coherence length and non-zero spatial overlap between successive meron bound states. This leads to noticeable discrepancies between contact currents that should be absent when the condensate recovers full interlayer coherence. To demonstrate this full recovery of interlayer coherence, we increase the interaction strength to $U=-7~eV$ thereby reducing the coherence length from 12 points in the $\hat{x}$-direction to 4 points and eliminating the spatial overlap between merons. In Fig. (\ref{fig:fullrec}), we plot the terminal current flow at each contact at the increased interlayer interaction strength and set $V_{TL} = 3~V$ to place the system firmly within the PFM-PPM oscillatory phase. Once again, the dips in the terminal current at times of $20~fs$ and $40~fs$ signal that a charged meron vortex has passed though the contact. By examining the terminal currents, we find that the current at bottom layer is fully recovered after vortex passes as the terminal currents have the following relation: $I_{TL}=-I_{BL}$, and $I_{TR}=-I_{BR}$.

\section{Charge of The Vortex State}
\begin{figure}
\includegraphics[width=0.45\textwidth]{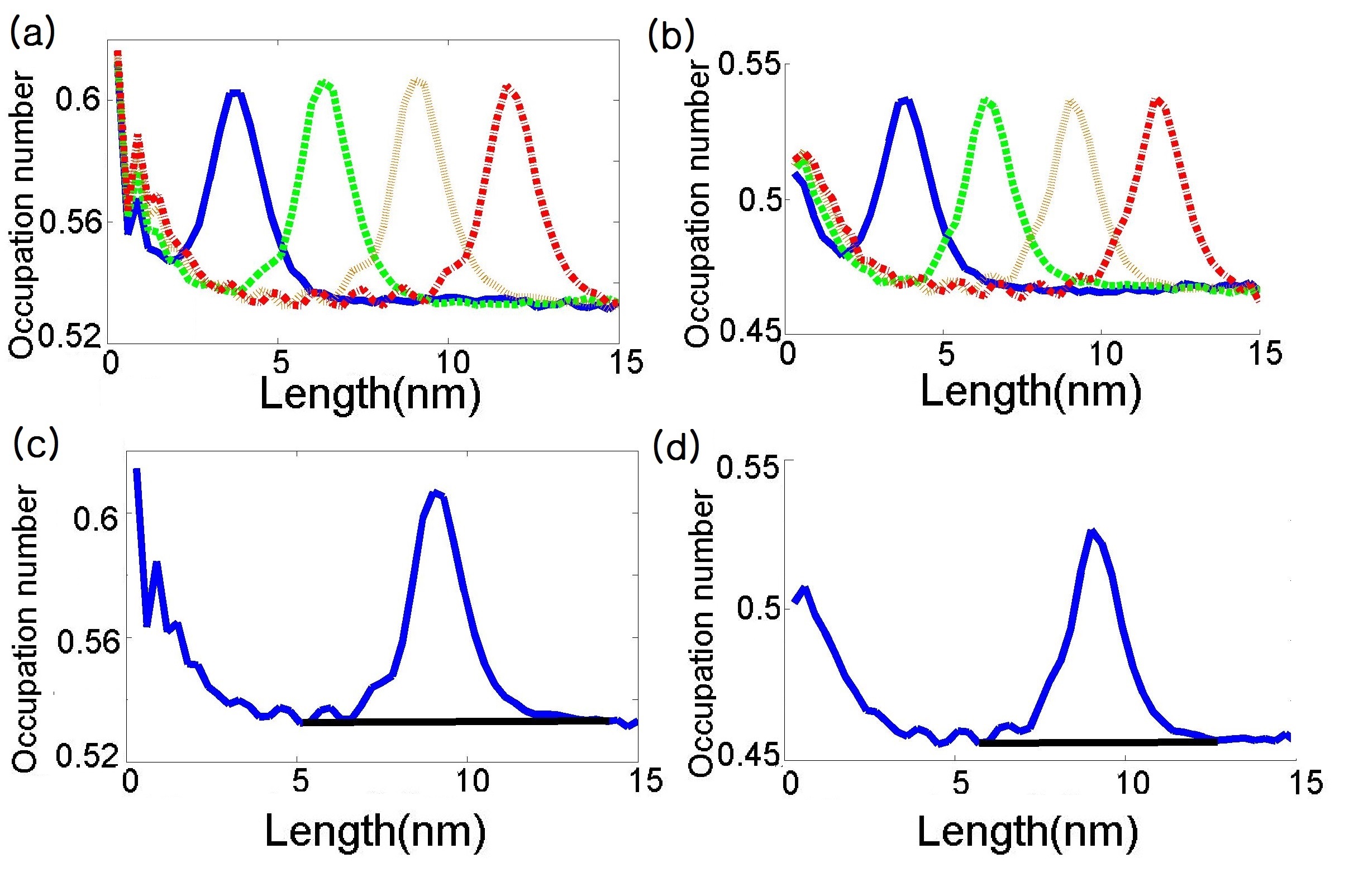}
\caption{\label{fig:vor} Time evolution of electron occupation number within the (a) top layer and (b) bottom layer. Each successive snapshot in the figure (blue, green, orange, and red) is taken at times of 13.8,15.8,17.8, and 19.7~fs respectively. Results of the integration scheme to obtain the relevant vortex charge within the (c) top layer and (d) bottom layer. The density profile is integrated with the background density (black solid line) at a time of 15.8~fs at an applied bias of $V_{TL}=3~V$.}
\end{figure}

In this section, we endeavor to clarify the charge of the vortex state within our system. The charge of vortex trapped state is numerically confirmed to be $\frac{q}{2}$ in each of the layer via the following procedure. At given time $t>0$, the density fluctuations within each layer are calculated from diagonal part of density matrix with the results plotted in Fig. (\ref{fig:vor}). For illustrative purposes, we have set the interlayer interaction strength to be $U=-7~eV$ thereby ensuring that the vortex states are not overlapping with one another by reducing the size of the numerical coherence length, which is calculated to be 4 lattice points. In order to best observe the vortex dynamics, we set $V_{TL}=3~V$ ensuring the persistent launching of charged merons associated with the metastable phase of the PFM-PPM. In Fig. \ref{fig:vor}(a) and Fig. \ref{fig:vor}(b), we observe the density fluctuations associated with the propagating merons centered at vortex core are propagating along with the current flow in the top and bottom layer, respectively. By numerically integrating density fluctuations and removing the background quasiparticle density, as shown in Fig. \ref{fig:vor}(c) and Fig. \ref{fig:vor}(d), we ascertain that our numerical method results in vortex charge of $q_{top}=0.49(5)e$ and $q_{bot}=0.49(1)e$ confirming that half electron charge localized within each layer.

\section{Frequency Dependence of Metastable PFM-PPM State}
In order to understand the frequency dependence of our results, we use the Gross-Pitaevskii equation,
\begin{equation}
 \label{eq:gpeq}
i\hbar \frac{\partial \psi_{ps}}{\partial t}=(-\frac{1}{2m}\nabla^2+V(r)+g|\psi_{ps}|^2)\psi_{ps}
\end{equation}
For a generic 1D system of length, $L$, we input our voltage profile of $V(0)=(V-V_c)$ and $V(L)=0~V$. From this we have
\begin{equation}
 \label{eq:gpeq_int}
 \frac{2e}{\hbar}(V-V_c)=\frac{\partial}{\partial t}(\arg(\psi(L))-\arg(\psi(0)))
\end{equation}
To obtain a stable condensate, there must be additional winding of the phase of the order parameter induced by the chemical potential difference across the system and the associated $\pi$ phase discontinuity of the order parameter in the merons. Therefore, we obtain the typical expression for the resultant AC Josephson frequency
\begin{equation}
 \label{eq:acfreq}
 f_{ps}=\frac{e}{h}(V-V_c)
\end{equation}

\begin{figure}
\includegraphics[width=0.45\textwidth]{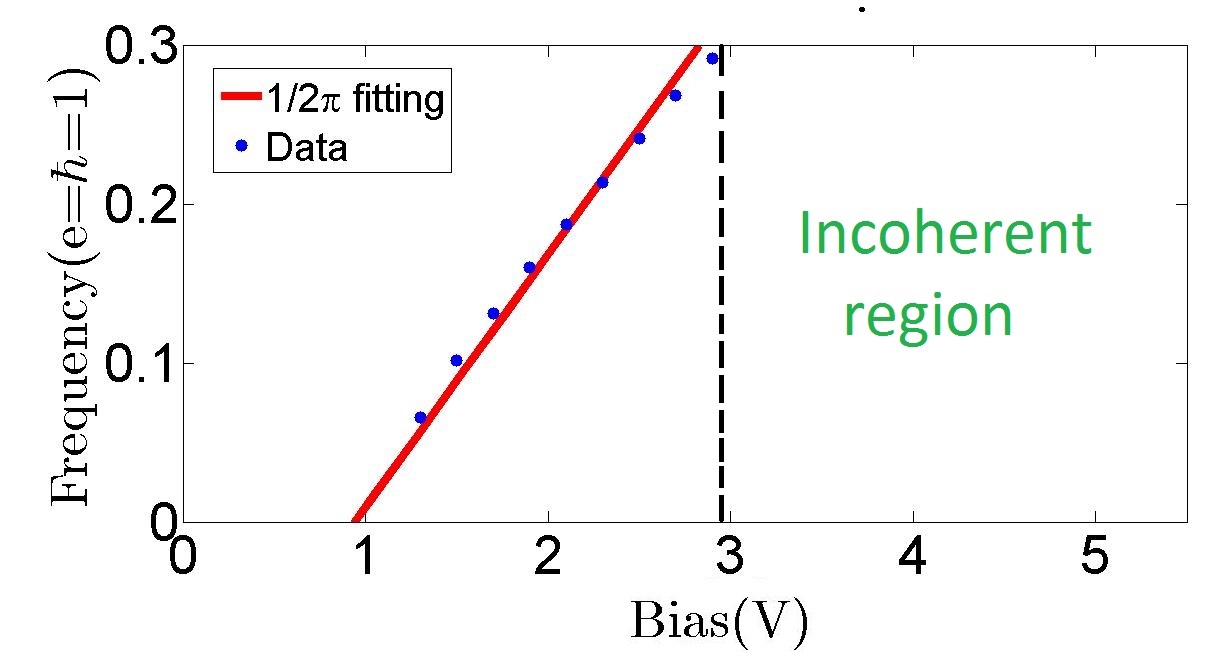}
\caption{\label{fig:fre} Plot of the voltage dependence of the oscillation frequency of the terminal currents. The red dashed line is $1/2\pi$ fitting while the points are data taken from numerical simulations of the DEC.  We observe an AC Josephson frequency dependence of the terminal currents past $I_c$.}
\end{figure}

In Fig. \ref{fig:fre}, we plot the frequency dependence of the terminal currents past $I_c$. We find the slope of the linear fitting to be $\frac{1}{2\pi}$ which confirms the vortex generation follows the AC Josephson frequency as predicted in Eq. (\ref{eq:acfreq}).  At low bias, which governs the linear portion of Fig. \ref{fig:fre}, we observe that as the interlayer voltage changes, the stable oscillations occur with increasing frequency. The critical voltage is naturally proportional to the interaction strength, $\Delta$, which is implicitly included in $V_c$ of Eq. (\ref{eq:acfreq}) as the strength of $U$ determines the location of metastable PFM-PPM phase boundary.

\section{Phase Boundary Dependence on the Interaction Strength}
\begin{figure}
\includegraphics[width=0.45\textwidth]{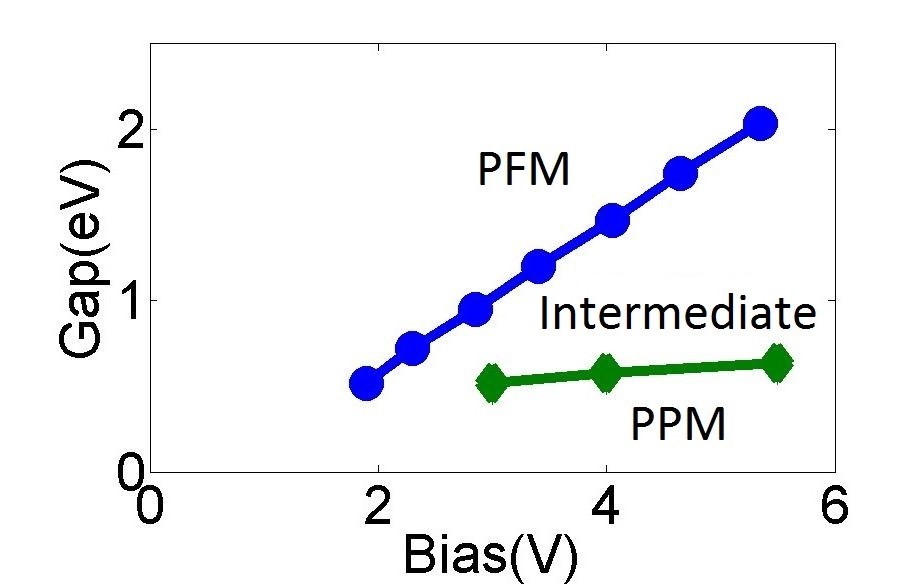}
\caption{\label{fig:udep} Phase boundaries of PFM-Metastable transition (blue circle) and Metastable-PPM transition (green diamond). The phases are calculated within a given time window of 10~fs. For each point of blue curve, we use interaction strengths of $U=-5,-5.5,-6,-6.5,-7,-7.5,-8~eV$ while, for green curve, $U=-5,-5.15,-5.3~eV$ are used. Strength of Hartree term is doubled to see the location of phase boundary in short time simulation.}
\end{figure}

Based on our results, it is clear that there is some dependence on the locations of the phase transitions and the strength of the interlayer interactions. We explore this relationship in Fig. \ref{fig:udep} which shows interaction strength dependence of phases. We find that the location of the phase boundary of PFM-PPM metastable transition is set by $V_c$ from Eq. (\ref{eq:acfreq}) in the Supplementary text. In Fig. \ref{fig:udep}, we find a clear linear dependence of $V_c$ with $\Delta$ when we examine the location of the phase transition between the PFM and PFM-PPM metastable phases. As the gap size increases with the increases in the interaction strength, the PFM phase stability to interlayer voltage increases accordingly with the critical voltage, $V_c$, moving to higher interlayer voltages. Additionally, we observe a similar trend in the transition between the metastable and PPM regions. In the zero gap limit, we know that the both the PFM and the PFM-PPM metastable phase must vanish. Therefore, in limit of infinite time response, intersection of the two boundaries must meet at origin of the plot. In Fig. \ref{fig:udep}, the intersection of the two lines is shifted from the origin due to nature of time dependent simulation. As we always have a finite time window within the simulation methodology associated with the TDKB formalism, it is inevitable to set a criteria of phase transition from given finite time simulation. This gives a time scale cutoff which shifts the phase boundaries from infinite time response limit. To be more precise, the Metastable-PPM transition is defined to be the point at which the interlayer coherence decreases to 30\% of initial self-consistently obtained value. Meanwhile, PFM-oscillation transition is defined to be a point where two merons are launched within 10~fs.

